\documentclass{elsarticle}
\usepackage[a4paper, top=2.5cm, left=2.5cm, bottom=2.5cm, right=2.5cm]{geometry}  

\journal{Solar Energy}

\usepackage{pgf}
\usepackage{amsmath,amssymb,bbm}

\biboptions{authoryear}

\newcommand{\eg}{\emph{e.g.}}
\newcommand{\ie}{\emph{i.e.}}
\newcommand{\SolTrack}{\emph{SolTrack}}
\newcommand{\mdeg}{^\circ}
\newcommand{\tdeg}{$^\circ$}

\newcommand{\tJd}{t_{\mathrm{Jd}}}
\newcommand{\tJc}{t_{\mathrm{Jc}}}

\usepackage[pdftex]{hyperref}
\hypersetup{
  colorlinks = true,  
  linkcolor = black,
  citecolor = black,
  urlcolor = black,
  pdftitle = SolTrack: a free fast and accurate routine to compute the position of the Sun,
  pdfauthor = Marc van der Sluys; Paul van Kan,
  pdfsubject = Solar Energy,
  pdfkeywords = Solar Energy Sun,
  pdfcreator = TeXLive LaTeX2e on Gentoo Linux,
  pdfproducer = MvdS - TeXLive LaTeX2e on Gentoo Linux
}

\begin{document}

\begin{frontmatter}
  
  \title{\SolTrack: a free, fast and accurate routine \\
    to compute the position of the Sun}
  
  
  \author[nikhefaddress,uuaddress]{Marc van der Sluys\corref{mycorrespondingauthor}}
  \cortext[mycorrespondingauthor]{Corresponding author}
  \ead{sluys@nikhef.nl}
  
  \author[hanaddress]{Paul van Kan}
  
  \address[nikhefaddress]{Nikhef, P.O.\ Box 41882, NL-1009 DB Amsterdam, The Netherlands.}
  
  \address[uuaddress]{Utrecht University, Institute for Gravitational and Subatomic Physics (GRASP), P.O.\ Box
    80000, NL-3508 TA Utrecht, The Netherlands.}
  
  \address[hanaddress]{HAN University of Applied Sciences, P.O.\,Box 2217, NL-6802 CE, Arnhem, The
    Netherlands.}

  \begin{abstract}
    We present a simple, free, fast and accurate C/C++ and Python routine called \SolTrack, which can compute
    the position of the Sun at any instant and any location on Earth.  The code allows tracking of the Sun
    using a low-specs embedded processor, such as a PLC or a microcontroller, and can be used for applications
    in the field of (highly) concentrated (photovoltaic) solar power ((H)CPV and CSP), such as tracking
    control and yield modelling.  \SolTrack\ is accurate, fast and open in its use, and compares favourably
    with similar algorithms that are currently available for solar tracking and modelling.  \SolTrack\
    computes $1.5 \times 10^6$ positions per second on a single 2.67\,GHz CPU core.  For the period between
    the years 2017 and 2116 the uncertainty in position is $0.0036 \pm 0.0042^\circ$, that in solar distance
    0.0017 $\pm$ 0.0029\%.  In addition, \SolTrack\ computes rise, transit and set times to an accuracy better
    than 1~second.  The code is freely available online.\footnote{\label{fn:url}http://soltrack.sf.net,
      https://pypi.org/project/soltrack/}
  \end{abstract}
  
  
\end{frontmatter}


\section{Introduction}

\noindent
For a project to develop affordable units for highly-concentrated photovoltaics (HCPV) in an urban
environment, we require an accurate ($\lesssim 0.01\mdeg$), fast and affordable algorithm that we can use
freely for commercial products.  Currently, existing codes in the field are either less accurate or have an
insufficiently open licence.  Since many codes that could potentially fulfil our demands can be found online,
albeit not currently in the field of solar concentration or in a preferred language (C/C++ and Python), we
considered a number of these codes and used them to develop the \SolTrack\ code.  Since many steps in any
algorithm to compute the position of the Sun can be performed in a number of ways, we selected the
alternatives that gave the speed-accuracy performance of our liking and assembled them to create \SolTrack.
We currently use our code to control our HCPV system and to model its electrical yield.  In
Section~\ref{sec:Descripion} we describe the origin and key features of the code that is included in
\SolTrack\ and the software licence under which it can be used.  In Section~\ref{sec:Performance} we present
the accuracy and computational speed of our code, and compare them to the two other algorithms that are
currently available in the field.  In Section~\ref{sec:Summary} we summarise our findings and present our
conclusions.

\section{Description of the \SolTrack\ code}
\label{sec:Descripion}

\noindent
The original \SolTrack\ code \citep{SolTrack} is written in plain C, which makes it available for many users
and computing platforms.  In addition, there is a Python version available.  It can compute the position of
the Sun in topocentric coordinates, for any geographical location on Earth and for any instant.  The code can
express the solar vector both in a horizontal (azimuth and altitude or zenith angle) and in a parallactic
coordinate system (using the equatorial coordinates hour angle and declination).  The \SolTrack\ code is
derived from the Fortran library \emph{libTheSky} \citep{libTheSky} and includes corrections for aberration
and parallax, and a simple routine to correct for atmospheric refraction \citep{refract}.  Apart from sky
position and distance, \SolTrack\ can compute rise, set and transit times for a given location and day, as
well as the corresponding azimuths and altitude.  The C/C++ code is freely available online at
\url{http://soltrack.sf.net}, under the terms and conditions of the GNU Lesser General Public Licence
\citep{GNU_LGPL}.  The Python code can be freely downloaded from the Python Package Index (PyPI) at
\url{https://pypi.org/project/soltrack/}.  A few examples that use \SolTrack\ are added to the download, so
that the users can verify their implementation.

The high accuracy, low computational cost and implementation in plain C and Python allow a flexible use of
\SolTrack, ensuring that the code can run on inexpensive, low-spec embedded systems like the STM32F4 DISCOVERY
boards, on simple systems with light-weight operating systems like the Raspberry Pi, on PLCs and on standard
PCs or servers.  A version ported to Arduino Scratch is available on the website.  In addition to \SolTrack,
we have developed a closed-loop system that allows feedback from sensors near the PV cell in order to correct
for discrepancies that arise due to \eg\ an imperfect installation of the system, or mechanical deformations.
Furthermore, we have designed an algorithm that takes into account non-perfect alignment, for example because
the system is aligned with the main axes of the building it resides in, rather than the north-south axis
\citep{CPV11_proceedings}.

\subsection{Calculation of the position of the Sun}

\noindent
The open-source Fortran library \emph{libTheSky} \citep{libTheSky} contains several routines to compute the
Sun's position with different performances in terms of accuracy and computational speed.  We used the
subroutine \texttt{sunpos\_la()} (for a low-accuracy Sun position\footnote{In fact, the library also provides
  routines for higher and lower accuracy than \texttt{sunpos\_la()}.}) as a development environment.  We used
the references in the code to read the literature behind it and see whether alternatives existed for each
equation.  Thus we were able to test the influence of the different parts of the code and their alternatives
on the accuracy and speed of the code.  In many cases, we simplified the original equations by leaving out
higher-order terms if the resulting increase in speed was significant and the loss in accuracy was limited or
negligible.  In some cases we found alternatives with a similar effect.  We translated the resulting code from
Fortran to C and Python as the basis for \SolTrack, in order to meet our own demands, as well as to make it
available to a large audience and a wide range of computing platforms.

By \emph{profiling} our code using gprof \citep[part of GNU binutils;][]{binutils} and Valgrind
\citep{valgrind} we found that the majority of the CPU cycles went to computing sines, cosines and tangents
and to the use of the \texttt{fmod()} function.  We reduced CPU time by computing the trigonometric functions
once and distributing the result through the code wherever that was possible.  Some angles, like the latitude,
declination and altitude, are defined in the range of $[-\frac{\pi}{2}, \frac{\pi}{2}]$
($[-90\mdeg, 90\mdeg]$) so that the cosine of an angle $\varphi$ can be computed cheaply from
$\cos\varphi = \sqrt{1 - \sin^2\varphi}$ without ending up in the wrong quadrant.  If all three of sine,
cosine and tangent of an angle must be computed, the latter can be cheaply obtained by division of the first
two.  In addition, we wrote our own \texttt{atan2()} function in C, which computes the inverse tangent without
confusion in the quadrants, based on \cite{wiki:atan2}.  If $\tan\alpha \equiv \frac{y}{x}$, then
\begin{equation}
  \alpha \equiv \operatorname{atan2}(y,x) =
  \begin{cases}
    \arctan(\frac y x)       & \text{if } x > 0;                      \\
    \arctan(\frac y x) + \pi & \text{if } x < 0 \text{ and } y \ge 0; \\
    \arctan(\frac y x) - \pi & \text{if } x < 0 \text{ and } y < 0;   \\
    +\frac{\pi}{2}           & \text{if } x = 0 \text{ and } y > 0;   \\
    -\frac{\pi}{2}           & \text{if } x = 0 \text{ and } y < 0;   \\
    0                        & \text{if } x = 0 \text{ and } y = 0,
  \end{cases}
  \label{eq:atan2}
\end{equation}
where in the last case, we return $0$ even though the result is undefined.  We found that this version is
about 39\% faster than the function provided in the C standard math library.  Finally, \SolTrack\ treats all
angles in radians, which avoids a computational overhead from constant conversion to and from degrees.  The
input and output can be easily converted from and to degrees at the beginning and end of the call
respectively.  A switch is available to the user to tell the routine that degrees are used for input and
output if wanted.

\subsubsection{Date, time and location}

\noindent
The desired date and time are provided as an input for the code in the common Gregorian calendar system.  As
is customary in astronomical calculations, these are converted to a Julian day \citep[JD; \eg][]{ESttAA_2012},
which provides a continuous time variable measured in days.  The JD is further converted to time in Julian
days ($\tJd$) and centuries ($\tJc$) since the year 2000.0, which are required for input in many of the
equations.  The geographic location of the observer is provided as the longitude $l_{\mathrm{obs}}$ (where
east from the Greenwich meridian is positive) and latitude $b_{\mathrm{obs}}$ (where the northern hemisphere
is positive), in either radians or degrees.

\subsubsection{Ecliptical coordinates}

\noindent
The ecliptic plane forms the base plane of the solar system, hence position calculations regarding orbital
motions of the planets are usually performed in ecliptical coordinates.  The geocentric position of the Sun is
easily computed from the heliocentric position of the Earth in its orbit around the Sun in ecliptical
coordinates by negating the ecliptical latitude and adding $\pi$ (180\tdeg) to the ecliptical longitude.

The ecliptical longitude is computed as follows.
The mean longitude \citep{Simon_ea_94} and mean anomaly \citep{ELP2000-85} of the Sun are given by (when
dropping unnecessary higher-order terms):
\begin{eqnarray}
  \lambda_0 &=& 4.895063168 + 628.331966786 \cdot \tJc + 5.291838 \times 10^{-6} \cdot \tJc^2; \\
  M &=& 6.240060141 + 628.301955152 \cdot \tJc  -  2.682571 \times 10^{-6} \cdot \tJc^2.
\end{eqnarray}
The Sun's equation of the centre is the difference between the true and mean anomalies and between the true
and mean longitudes of the Sun.  It can be approximated using \citet[][up to the $\sin 2M$ term]{Meeus98}:
\begin{eqnarray}
  C &=& \left(3.34161088 \times 10^{-2} - 8.40725 \times 10^{-5}  \cdot \tJc - 2.443 \times 10^{-7} \cdot \tJc^2\right) \sin M \nonumber \\
    &+& \left(3.489437 \times 10^{-4} - 1.76278 \times 10^{-6} \cdot \tJc\right) \sin 2M.
\end{eqnarray}
The true longitude is then simply:
\begin{equation}
  \lambda = \lambda_0 + C.
\end{equation}

In order to apply a correction to the Sun's position for nutation and aberration, we need the longitude of the
\emph{Moon}'s mean ascending node $\Omega$ \citep[][up to second order]{ELP2000-85}:
\begin{equation}
  \Omega = 2.1824390725 - 33.7570464271 \cdot \tJc + 3.622256 \times 10^{-5} \cdot \tJc^2.
\end{equation}
The \emph{nutation in longitude} is a periodic wobble of the Earth's rotation axis due to the presence of the
Moon and can be described with sufficient accuracy by \citep[][leading term only]{IAU1980nutation}:
\begin{equation}
  \Delta\psi = -8.338601 \times 10^{-5} \sin\Omega.
\end{equation}

If an accurate distance is important, we compute the geocentric distance of the Sun in AU from the
eccentricity of the Earth's orbit $e$ \citep[][to second order]{Simon_ea_94} and the true anomaly $\nu$:
\begin{eqnarray}
  e &=& 0.016708634 - 4.2037 \times 10^{-5} \cdot \tJc  -  1.267 \times 10^{-7} \cdot \tJc^2; \\
  \nu &=& M + C; \\
  R &=& 1.0000010178 ~ \frac{1 - e^2}{1 + e \cos\nu}.
\end{eqnarray}
In other cases, we simply assume a circular orbit:
\begin{equation}
  R = 1.0000010178.
\end{equation}
\emph{Annual aberration} is the effect that the light from a celestial body seems to come slightly more from
direction of the Earth's motion around the Sun, similar to the apparent motion of rain drops coming from the
direction a car is moving into (but much less so, since the orbital speed of the Earth is much smaller than
the speed of light).  We use \citep{fuas.book.04}:
\begin{equation}
  \Delta\lambda = \frac{-9.93087 \times 10^{-5}}{R}.
\end{equation}

We can now correct the longitude of the Sun for nutation and aberration, and arrive at the ecliptical
longitude at the given instant:
\begin{equation}
\lambda = \lambda + \Delta\lambda + \Delta\psi.
\end{equation}

Since we do not need extremely high accuracy, and since the Sun is (by definition) always near the ecliptic,
for the ecliptical latitude of the Sun we simply use
\begin{equation}
\beta = 0.
\end{equation}

Finally, we need the obliquity of the ecliptic $\varepsilon$, the angle between the Earth's rotation axis and
the normal to the ecliptic plane, for coordinate transformations later on.  However, we calculate it here
because $\Omega$ is now available.  We compute it from the mean obliquity $\varepsilon_0$ \citep{Meeus98} and
the nutation in obliquity $\Delta\varepsilon$ \citep[leading term only]{IAU1980nutation}:
\begin{eqnarray}
  \varepsilon_0 &=& 0.409092804222 - 2.26965525 \times 10^{-4} \cdot \tJc - 2.86 \times 10^{-9} \cdot \tJc^2; \\
  \Delta\varepsilon &=& 4.4615 \times 10^{-5} \cos\Omega; \\
  \varepsilon &=& \varepsilon_0 + \Delta\varepsilon.
\end{eqnarray}

\subsubsection{Equatorial and parallactic coordinates}

\noindent
The equatorial coordinates \emph{right ascension} $\alpha$ and \emph{declination} $\delta$ are computed from
the ecliptical coordinates $(\lambda,\beta)$ using a standard coordinate transformation
\citep[\eg][]{ESttAA_2012}:
\begin{eqnarray}
  \tan\alpha &=& \frac{\sin\lambda \, \cos\varepsilon - \tan\beta \, \sin\varepsilon}{\cos\lambda}; \label{eq:ra} \\
  \sin\delta &=& \sin\beta \, \cos\varepsilon + \cos\beta \, \sin\varepsilon \, \sin\lambda,
\end{eqnarray}
where we use the \texttt{atan2()} function to compute the right ascension from Eq.\,\ref{eq:ra}.

The parallactic coordinate hour angle is the difference between the local sidereal time $\theta$ and the right
ascension:
\begin{equation}
  H = \theta - \alpha,
\end{equation}
where
\begin{eqnarray}
  \theta       &=& \theta_0 + \Delta\theta + l_{\mathrm{obs}}; \\
  \theta_0     &=& 4.89496121 + 6.300388098985 \cdot \tJd + 6.77 \times 10^{-6} \cdot \tJc^2;  \\ 
  \Delta\theta &=& \Delta\psi \cdot \cos\varepsilon.                                           
\end{eqnarray}
Here, $\theta$ is the local sidereal time, $l_{\mathrm{obs}}$ is the observer's geographical longitude (east
of Greenwich is positive), $\theta_0$ is the Greenwich mean sidereal time and $\Delta\theta$ corrects for
nutation in right ascension, converting to \emph{apparent} Greenwich sidereal time.

Together, hour angle and declination provide the coordinates needed for a \emph{parallactic} mount.

\subsubsection{Horizontal coordinates}

\noindent
We use the parallactic coordinates hour angle $H$ and declination $\delta$ to compute the horizontal
coordinates \emph{azimuth} $A$ and \emph{altitude} $h$.  This too is a well known coordinate transformation
\citep[\eg][]{ESttAA_2012}:
\begin{eqnarray}
\sin h &=& \sin\delta \, \sin b_{\mathrm{obs}} +  \cos H \, \cos\delta \, \cos b_{\mathrm{obs}}; \label{eq:altitude} \\
\tan A &=& \frac{\sin H}{\cos H \, \sin b_{\mathrm{obs}} - \tan \delta \, \cos b_{\mathrm{obs}}}, \label{eq:azimuth}
\end{eqnarray}
where $b_{\mathrm{obs}}$ is the geographic latitude of the observer (northern hemisphere is positive).  We use
the \texttt{atan2()} function again to compute the azimuth from Eq.\,\ref{eq:azimuth}.

A variable that is often used instead of the altitude is the \emph{zenith angle}, which is simply given by
\begin{equation}
  \zeta = \frac{\pi}{2} - h.
\end{equation}

\subsubsection{Correction for atmospheric refraction}

\noindent
On the last part of their journey to the Earth's surface, the Sun's rays travel through the atmosphere of the
Earth.  The effect of the increasing density of the atmosphere causes refraction of the light rays towards the
Earth's surface, so that they generally hit the surface somewhat earlier (nearer the Sun) than if the
atmosphere had been absent.  The result for an observer on the ground is that the altitude of the Sun $h$
appears to be somewhat higher than we have just computed.  The effect vanishes if the Sun is in the zenith,
and has a maximum close to the horizon of about 0.5\tdeg, roughly the Sun's apparent diameter.  For accurate
positions, we need to take this effect into account, and because atmospheric refraction affects the altitude
only, this is most easily done in \emph{horizontal} coordinates.

We use a simple prescription by \cite{refract}, which, when expressed in radians, looks as follows:
\begin{equation}
  \Delta h = 0.0002967 \cot\left(h + \frac{0.0031376}{h + 0.0892}\right) ~
  \left(\frac{T}{283}\right)^{-1} \left(\frac{P}{101}\right).
\end{equation}
Here, $\cot()$ is the cotangent, $h$ the uncorrected altitude, $T$ the temperature in Kelvin and $P$ the
atmospheric pressure in kPa.  For standard atmospheric values of $P = 101$\,kPa and $T=283$\,K (10\tdeg C),
the last two fractions can be ignored.  We then \emph{add} the quantity $\Delta h$ to the uncorrected altitude
$h$ to account for the atmospheric refraction.  Alternatively, we \emph{subtract} $\Delta h$ from the
uncorrected zenith angle $\zeta$.

\SolTrack\ also offers refraction-corrected \emph{parallactic} coordinates.  These are obtained by
transforming the corrected horizontal coordinates back to the parallactic system using the inverse
transformation of Eqs.\,\ref{eq:altitude}--\ref{eq:azimuth}.

\section{Performance of the \SolTrack\ code}
\label{sec:Performance}

\subsection{Accuracy of \SolTrack}
\label{sec:accuracy}
\noindent
In order to assess the accuracy of our code, we need a comparison.  We used the VSOP\,87 code to compute very
precise positions of the Sun, with an accuracy of $1.4 \times 10^{-6}$ between the years 1900 and 2100
\citep{VSOP87}.  In order to correct for atmospheric refraction, we used an accurate model
\citep{accurRefract} that integrates the path of a light beam as it travels through the atmosphere for a given
\emph{apparent} sky position.  This model is used in a slow, iterative procedure to solve the inverse problem,
in order to find the apparent position for a given \emph{true} sky position, as implemented in
\emph{libTheSky} \citep{libTheSky}.  We assumed standard atmospheric pressure and temperature.  The accuracy
of the result will be limited by the detailed model for atmospheric refraction.  In this section, we assume
that the results from the combination of these two accurate models are exact, so that any deviation between it
and \SolTrack\ indicates an inaccuracy in our code.  This assumption is validated by the fact that the
detailed models are about three orders of magnitude more accurate than \SolTrack.

The comparison between VSOP\,87 and \SolTrack\ was carried out for 100,000 random moments in the next 100
years (between 2017 and 2116) when the Sun is above the horizon in the Netherlands. We find that the error in
position (either in parallactic or horizontal coordinates) is $0.0036 \pm 0.0042^\circ$, or about 0.68\% of
the apparent diameter of the Sun.  This is sufficient for solar tracking of HCPV systems under all conditions.

\begin{figure}
  \centering
  \pgfimage[interpolate=true, width=0.8\textwidth]{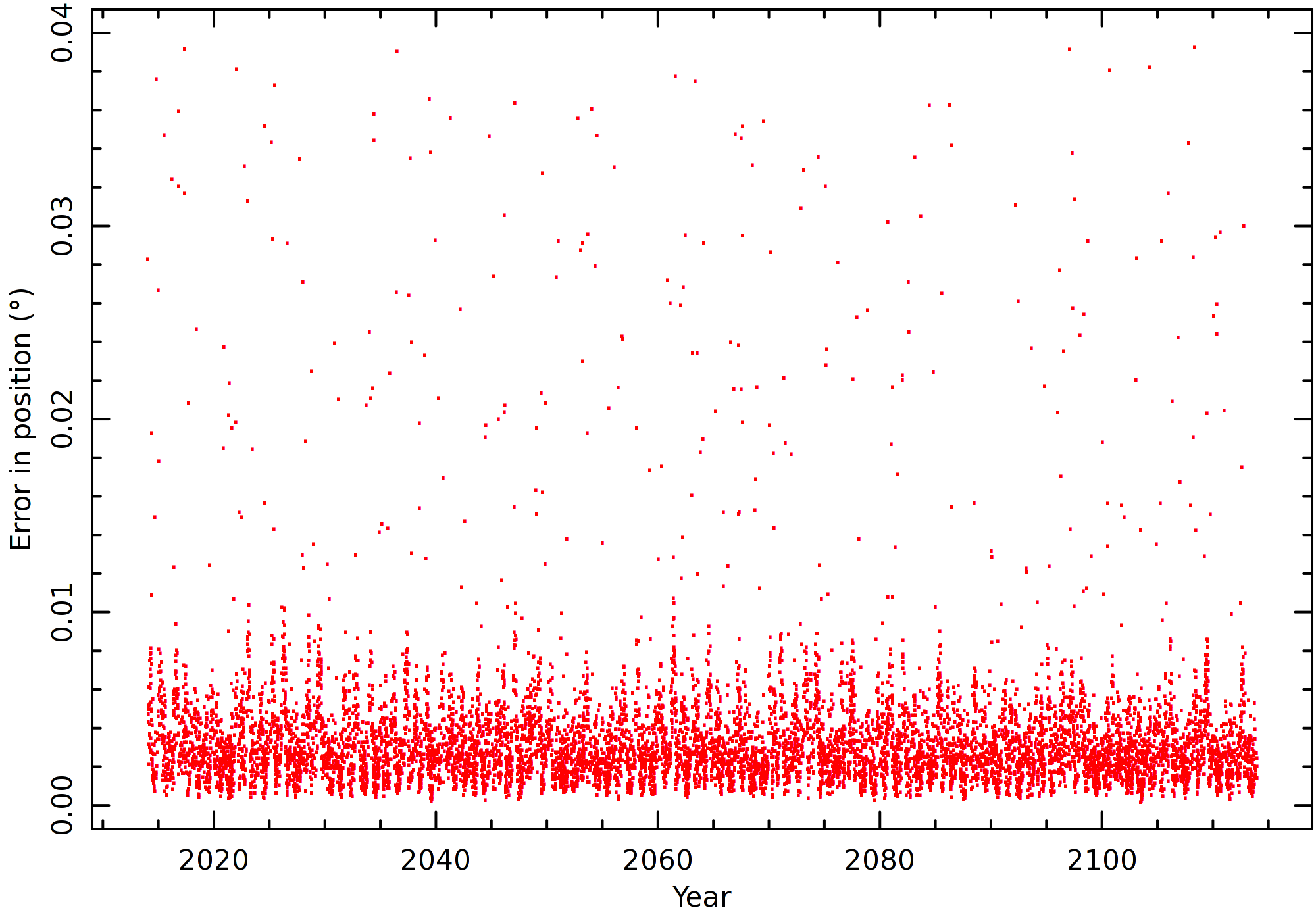}
  \caption{
    The accuracy of the \SolTrack\ algorithm as a function of time for 10,000 
    random instances where the Sun is above the horizon in Arnhem, the Netherlands
    between 2017 and 2116. While for most cases, the accuracy is better than 
    0.01\tdeg, a relatively small number of data points indicate a larger 
    deviation.  All these are near sunrise or sunset (see Fig.\ref{fig:STaccur}).
    \label{fig:STaccurYear}
  }
\end{figure}

Figure~\ref{fig:STaccurYear} shows the positional error for 10,000 of the computed sky positions as a function
of time over the next 100 years.  We see that the accuracy hardly deteriorates over the coming century, so
that the \SolTrack\ algorithm can be used safely for at least the next 100 years.  The figure also shows that
most positions yield an accuracy that is better than 0.01\tdeg, while only a small fraction of points are less
accurate than that.  The latter group represents very low Sun positions, $\lesssim 2.5\mdeg$, where the
correction for atmospheric refraction becomes less accurate.  If these low-altitude points are ignored, on the
grounds that very little energy will be generated at such Sun altitudes, the accuracy, and in particular its
standard deviation, drop to $0.0030 \pm 0.0016\mdeg$.

The accuracy of the computed distance is 0.0017\% $\pm$ 0.0029\%, which results in an typical error in the
computed solar power of 0.0058\%.  This will be completely negligible in a model where the uncertainty in
weather conditions will have a much larger impact.  The accuracy of the rise, transit and set times has been
determined to be better than one second, while the corresponding azimuths and altitudes have an accuracy that
is better than $10^{-3\circ}$.  The first uncertainty is far smaller than needed, since weather conditions
provide a much larger source of uncertainty.\footnote{Typically, the weather imposes an uncertainty of at
  least one minute on rise and set times.}  The second uncertainty is better than that for the position in
general (since it is a one-dimensional problem, rather than a two-dimensional one).

\subsubsection{Comparison of \SolTrack's accuracy to other algorithms}
\begin{figure}
  \centering
  \pgfimage[interpolate=true, width=0.8\textwidth]{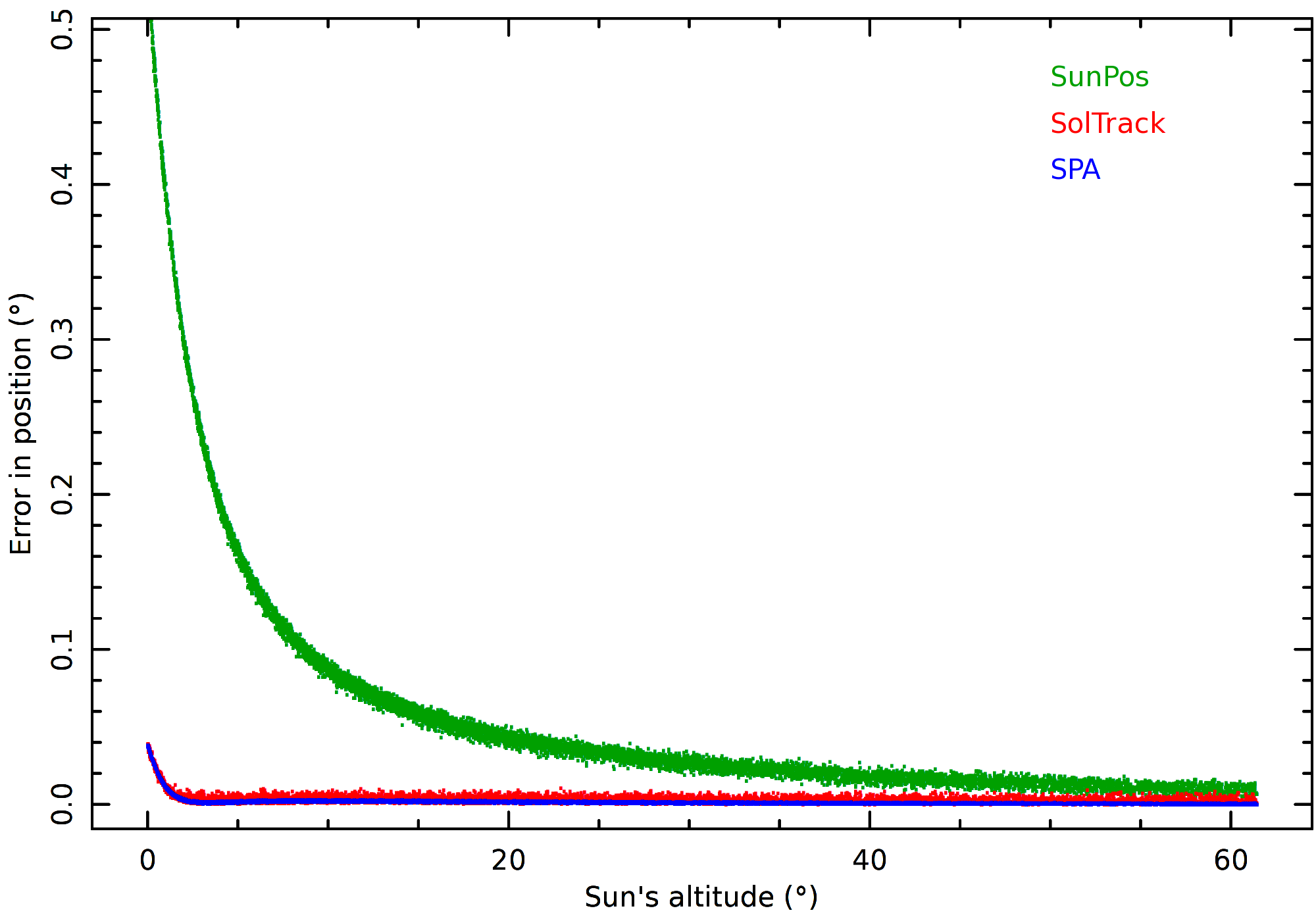}
  \caption{A comparison of the accuracies of PSA's \emph{SunPos} (green/grey, at the top), \SolTrack\
    (red/grey, near the bottom) and SPA (blue/black, at the bottom), as a function of altitude for 10,000
    random instances where the Sun is above the horizon in Arnhem, the Netherlands.  The data points for
    \SolTrack\ partially overlap with those for SPA (see Fig.\,\ref{fig:STaccurHist} for a clearer view).
    The largest difference is found between \emph{SunPos} and the other two codes, especially for low
    altitudes, mainly due to the lack of correction for atmospheric refraction in \emph{SunPos}.
    \label{fig:STaccur}
  }
\end{figure}

\begin{figure}
  \centering
  \pgfimage[interpolate=true, width=0.7\textwidth]{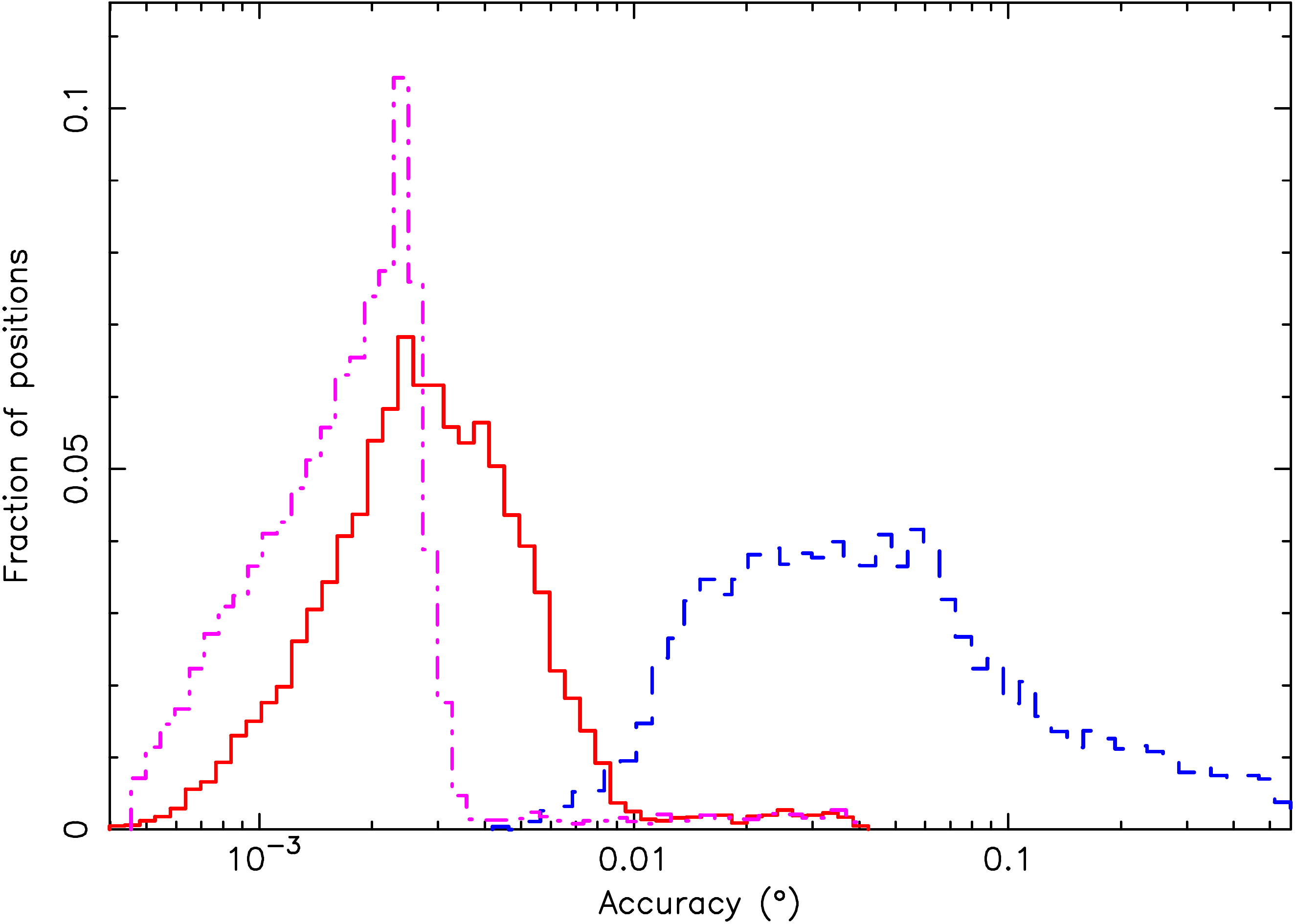}
  \caption{Three histograms comparing the accuracies of SPA (dash-dotted/purple line, to the left),
    \SolTrack\ (solid/red line, in the middle) and PSA's \emph{SunPos} (dashed/blue line, to the right),
    for 10,000 random instances where the Sun is above the horizon in Arnhem, the Netherlands.
    Note the logarithmic scale on the horizontal axis.
    \label{fig:STaccurHist}
  }
\end{figure}

\noindent
We have computed sky positions for the Sun for the same 100,000 random moments in next 100 years using PSA's
\emph{SunPos} \citep{PSA} and NREL's SPA \citep{SPA}, and compared them to the VSOP87 positions.  The results
are displayed in Figures~\ref{fig:STaccur} and \ref{fig:STaccurHist} and in Table~\ref{tab:compare}.

When compared to \emph{SunPos} routine, which is also lightweight and freely available, \SolTrack\ is about 20
times more accurate in both horizontal and parallactic coordinates.  Compared to the performance of the code
to the NREL routine SPA, which is more elaborate and has a more restricted licence, SPA is 36\% more accurate
than \SolTrack\ when comparing horizontal coordinates.  However, SPA does not offer refraction-corrected
parallactic coordinates, so that our code is about 20 times more accurate when a parallactic mount is used.
For this comparison we computed only the Sun's position in SPA, no rise and set times or incident radiation.
An overview of the comparison between the three codes can be found in Table~\ref{tab:compare}.

\begin{table}
  \centering
  \caption{
    Comparison of the accuracy, resulting loss of power w.r.t.\ optimal tracking 
    with and without secondary optics (SO)
    between the \emph{SunPos}, SPA and \SolTrack\ routines.  
    \label{tab:compare}
  }
  
  \begin{tabular}{lrccc}
    \hline
                             &             & \textbf{\emph{SunPos}} & \textbf{SPA} & \textbf{\SolTrack} \\
                             &             &                        &              &                     \\
    \hline
    \textbf{Accuracy}        & mean        & 0.073\tdeg             & 0.0023\tdeg  & 0.0036\tdeg \\
    ~~~\textbf{in position}  & st.dev.\    & 0.091\tdeg             & 0.0036\tdeg  & 0.0042\tdeg \\
                             & relative    & 20.2                   & 0.64         & 1.00 \\
    
    \multicolumn{5}{}{} \\
    \textbf{Accuracy}        & mean        & ---                    & 0.0017\%     & 0.0029\% \\
    ~~~\textbf{in distance}  & st.dev.\    & ---                    & 0.0011\%     & 0.0021\% \\
    ~~~\textbf{in power}     & mean        & ---                    & 0.0034\%     & 0.0058\% \\
    \hline
  \end{tabular}
\end{table}

\subsection{Speed of \SolTrack}
\noindent
In order to benchmark the \SolTrack\ C code, we timed runs where $10^7$ (ten million) positions were computed
on a single 2.67\,GHz CPU core of a normal laptop computer.  We used the \texttt{GCC} compiler
\citep[v6.2][]{gcc} with \texttt{-O2} optimisation and the \texttt{taskset} Linux utility to force the program
to always run on the same core \citep{util-linux}.  In these speed-benchmark runs we generated the date and
time randomly, in order to avoid an I/O overhead from reading these data from disc.  We did not save the
computed results for the same reason.

We performed ten such runs and found a mean run time and standard deviation (averaged over the ten runs of
$10^7$ calls to the \SolTrack\ routine each) of 7.331 $\pm$ 0.024\,s.  We then repeated the test, but without
actually calling the \SolTrack\ routine, which took 0.860 $\pm$ 0.012\,s (again the mean of ten runs of $10^7$
calls each).  This number represents the overhead from the code that \emph{calls} \SolTrack\ rather than from
\SolTrack\ itself, \eg\ from starting the code and generating the random date and time.  Hence, we determine
the CPU time for $10^7$ calculations of the Sun's position on the used CPU core as the difference of these two
numbers, \ie\ 6.471 $\pm$ 0.027\,s, or about $1.5 \times 10^6$ calls per second.\footnote{Scaled with the
  clock speed of the core, a 1.7\,kHz CPU would be needed in order to compute the Sun's position every second
  if this were its only task.}  These numbers are valid for the default mode of \SolTrack\, where the
refraction correction is computed only for horizontal coordinates and the Sun-Earth distance is not computed.
In these tests we only computed positions of the Sun; we did not determine rise, transit or set times.

In addition, we computed the same numbers, \ie\ the mean CPU time for ten runs of $10^7$ \SolTrack\ calls
each, corrected for the overhead determined from runs without calls to \SolTrack\, for the other three modes
of \SolTrack\: where the distance is computed as well and/or the equatorial coordinates are corrected for
atmospheric refraction as well.  The results are shown in Table~\ref{tab:speed_st}.  We find that computing
the distance adds about 8\% to the CPU time, while correcting equatorial coordinates for refraction costs 26\%
extra.

\begin{table}
  \centering
  \caption{
    Comparison of the CPU times for the calculation of $10^7$ positions between 
    modes where neither the distance or refraction correction for equatorial 
    coordinates are computed, or where one or both are computed.
    \label{tab:speed_st}
  }
  
  \begin{tabular}{lrcccc}
    \hline
    \textbf{Distance?}     &          & $\times$  & \checkmark  & $\times$    & \checkmark  \\
    \textbf{Ref corr eq.?} &          & $\times$  & $\times$    & \checkmark  & \checkmark  \\
    \hline
    
    \textbf{CPU time}   & mean        & 6.471\,s  & 6.987\,s    & 8.160\,s    & 8.686\,s \\
                        & st.dev.\    & 0.027\,s  & 0.028\,s    & 0.022\,s    & 0.030\,s \\
                        & relative    & 1.000     & 1.080       & 1.261       & 1.342 \\
    \hline
  \end{tabular}
\end{table}

\subsubsection{Comparison of \SolTrack's computational speed to other algorithms}
\noindent
We compare the CPU times of \SolTrack\ to those of \emph{SunPos} \citep{PSA} and SPA \citep{SPA} in
Table~\ref{tab:speed_comp}.  The table shows that \SolTrack\ is about 2.5\% faster than \emph{SunPos} if the
refraction correction is only applied to horizontal coordinates, and 23\% slower if this correction is also
applied for equatorial coordinates.  Hence, \SolTrack's performance in computational speed is similar to that
of \emph{SunPos}, despite the much greater accuracy of our code (see Sect.\,\ref{sec:accuracy}).

\begin{table}
  \centering
  \caption{
    Comparison of the CPU times for the calculation of $10^7$ positions between 
    \emph{SunPos}, SPA and \SolTrack\ routines.  The last column shows the results
    for \SolTrack\ with the additional refraction correction in equatorial 
    coordinates.
    \label{tab:speed_comp}
  }
  
  \begin{tabular}{lrcccc}
    \hline
                    &             & \textbf{\emph{SunPos}} & \textbf{SPA} & \textbf{\SolTrack} & \textbf{\SolTrack\ + eq.} \\
    \hline
    
    \textbf{CPU}    & mean        & 6.632\,s & 160.78\,s & 6.471\,s & 8.160\,s \\
    \textbf{time}   & st.dev.\    & 0.026\,s & 0.45\,s   & 0.027\,s & 0.022\,s \\
                    & relative    & 1.025    & 19.70     & 1.000    & 1.261    \\
    \hline
  \end{tabular}
\end{table}

When comparing \SolTrack\ to NREL's PSA code, our code is almost a factor of 20 faster if the refraction
correction is limited to horizontal coordinates, and nearly 16 times faster if \SolTrack\ corrects equatorial
coordinates for refraction as well, an option that does not exist in PSA.  As we saw in the previous section,
PSA gives more accurate results for its higher computational costs, albeit not by more than an order of
magnitude.

\section{Summary and conclusions}
\label{sec:Summary}

\noindent
In this paper, we presented \SolTrack, a free, fast and accurate code to compute the position of the Sun and
do related calculations.  Due to its open-source licence and the fact that the code is written in plain C and
Python, it can be used by a wide range of users on a wide range of computing platforms and for a wide range of
purposes, be it scientific, commercial or otherwise.  The mean accuracy over the period 2017--2116 is 0.0036
$\pm$ 0.0042\tdeg, and if instances where the Sun is less than 2.5\tdeg\ above the horizon are omitted, this
improves to 0.0030 $\pm$ 0.0016\tdeg.  \SolTrack\ can also compute rise, transit and set times to an accuracy
better than 1~second.  \SolTrack\ can compute millions of positions per second on a typical modern CPU, and as
such is capable of accurately tracking a solar-concentration system on a low-spec platform, as well as perform
many calculations in a yield model (\eg\ using a Monte Carlo method).

When compared to the two codes that are currently used in the field of (high-)concentration photovoltaics
((H)CPV) and concentrated solar power (CSP), it is comparable in its open licence and computing speed to the
first (PSA) while being about 20 times more accurate, and only slightly less accurate than the other (SPA)
whilst about 20 times faster and much freer in its licence.


\bibliography{SolTrack}

\end{document}